\documentclass[conference,compsoc]{IEEEtran}
\IEEEoverridecommandlockouts
\usepackage[nocompress]{cite}
\usepackage{amsmath,amssymb,amsfonts}
\usepackage[ruled]{algorithm}
\usepackage{algpseudocode}
\usepackage{graphicx}
\usepackage{textcomp}
\usepackage[table]{xcolor}
\usepackage{pifont}
\usepackage{hyperref}
\usepackage{cleveref}
\usepackage{xspace}
\usepackage{tcolorbox}
\usepackage{colortbl}
\usepackage{tabularx}
\usepackage{array}
\usepackage{amsthm}
\usepackage{enumitem}
\usepackage{multirow}
\usepackage{arydshln}
\usepackage{booktabs}
\usepackage{ragged2e}
\usepackage{makecell}
\usepackage{pbalance}
\def\BibTeX{{\rm B\kern-.05em{\sc i\kern-.025em b}\kern-.08em
    T\kern-.1667em\lower.7ex\hbox{E}\kern-.125emX}}

\begin{document}

\newcommand{\sysname}{\text{ReproCue}\xspace}
\newcommand{\code}[1]{{\tt\scriptsize #1}}
\newcommand{\codeztext}[1]{{\tt\footnotesize #1}}
\title{Reproducer Synthesis for AI-Reported Bugs in the Linux Kernel}

\newcommand{\paperperson}[3]{\makebox[0.33\textwidth]{\begin{tabular}[t]{@{}c@{}} {\large #1}\\
      {\small\itshape #2}\\
      {\small #3}
    \end{tabular}}}

\author{\normalfont\normalsize
  \begin{tabular}{@{}c@{}}
    \paperperson{Xiao Wu}
      {The Chinese University of Hong Kong}{xwu@cse.cuhk.edu.hk}\paperperson{Yunbo Ni}
      {The Chinese University of Hong Kong}{ybni@cse.cuhk.edu.hk}\paperperson{Hao Sun}
      {ETH Zurich}{hao.sun@inf.ethz.ch}\\
    \noalign{\vskip 0.9em}
    \paperperson{Yuanyuan Yuan}
      {Tsinghua University}{yyyuan@tsinghua.edu.cn}\paperperson{Shaohua Li}
      {The Chinese University of Hong Kong}{shaohuali@cuhk.edu.hk}
  \end{tabular}}

\maketitle

\begin{abstract}
Agentic review systems are increasingly flagging suspected bugs in Linux kernel patches as textual reports. A maintainer who receives such reports needs further inspection to determine whether the report is true. The inspection requires understanding whether the bug can be triggered and, if necessary, manually constructing one. As a consequence, despite the excessive textual explanation, many reports go unanswered. In this paper, we argue that a runnable proof-of-concept (PoC) is the key information in such AI-generated kernel review reports. It is, however, challenging to automatically synthesize a valid and runnable PoC.

We present \sysname, a framework that generates PoCs for AI-generated Linux kernel reviews. \sysname replays the historical kernel bugs and learns one reproduction fininte-state-machine (FSM) per subsystem from successful runs. An FSM records the steps a reproduction goes through, together with the requirements observed at each step. Given a new review, \sysname follows the FSM of that subsystem, adjusting the PoC and the environment as execution shows what is missing. In 100 recent AI-generated reviews, \sysname writes PoCs for 49 of them, compared with 10 for the current agentic reproducer and 9 for the state-of-the-art directed fuzzing. Excitingly, with our generated PoCs, Linux kernel maintainers confirmed 69.4\% of them, much higher than 9.3\% from Sashiko where only textual explanations are included, making AI reviews more actionable in practice.

\end{abstract}

\IEEEpeerreviewmaketitle

\section{Introduction}

AI agents are increasingly used in Linux kernel development~\cite{kernel_ai_assistants_doc}.
They write patches and review code changes for current kernel development. However, a large fraction of submitted patches remain under review, and CVE assignments have grown so rapidly that maintainers can only examine a small subset in depth~\cite{nist_nvd_2026}.
AI review tools have emerged to assist kernel maintainers. For example, Google’s Sashiko~\cite{sashiko} inspects new kernel patches, and the bpf-next bot~\cite{bpf-agent} triages BPF patch series in CI pipelines. Yet the workload of kernel maintainers continues to grow,
the burden on a stagnant pool of human maintainers, who must ultimately judge every flagged defect, often without executable PoC programs, is heavier than ever, with 55.29\%~\cite{lin2026learning} kernel patches receiving no response over the past year.

\begin{figure}
    \centering
    \includegraphics[width=0.48\textwidth]{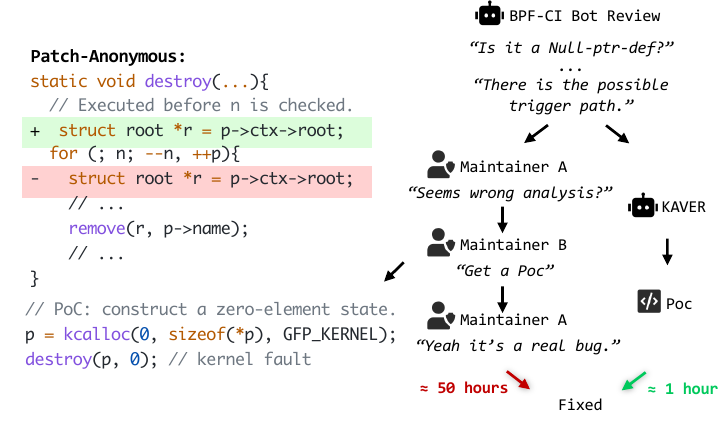}
    \caption{A motivating kernel discussion demonstrating the practical value of PoC-backed AI review. A reproducing PoC resolved a disputed warning and provided executable evidence that the reported defect was reachable.
    }
    \label{fig:intro_example}
\end{figure}

Current AI review tools stop at textual bug reports~\cite{sashiko,bpf-agent}. As a result, maintainers still have to verify whether each reported failure is real and reproducible.
The limited maintainer response and imperfect review correctness both highlight this burden. For example, Google's Sashiko review emails receive replies in only 21.2\% of cases, while an evaluation of 48 review threads found 16 incorrect reviews~\cite{bpf-agent-review}. Without an executable PoC, maintainers must validate the reported failures from the textual analysis alone.
We argue that a capable review agent should ideally search for a runnable PoC, which can significantly improve the confidence level of the report and reduce maintainers' burden.

Unfortunately, existing techniques do not adequately close this review-to-PoC gap.
Although fuzzing can target specific code locations or crash classes~\cite{syzkaller2024,tan2023syzdirect}, limitations in path and state coverage can hinder reproduction of the particular bug identified in a review.
General-purpose agents, e.g., Claude Code~\cite{ClaudeCode}, lack a governed kernel build and VM feedback loop~\cite{arora2025setupbench}, while dedicated agentic reproducers such as K-Repro~\cite{krepro} assume a largely fixed environment.
Harnesses that are too flexible or too rigid cannot meet the needs of exploration and exploitation in the Linux kernel~\cite{krepro}. They cannot reliably adapt architectures, kernel configurations, devices, resources, and userspace tools as their requirements emerge during execution.

\Cref{fig:intro_example} shows an example of how a PoC can resolve a disputed AI review.
On the left side, the patch moved the dereference of \codeztext{p->ctx->root} from inside a count-controlled loop to before it. Consequently, when the element count is zero, the new code dereferences \codeztext{p} even though the loop itself is skipped. A follow-up PoC constructed this zero-element state, providing the executable evidence that led the developer to acknowledge and revise the defective patch. On the right side, maintainers shifted from initial AI skepticism to acceptance upon receiving PoC, and finally fixed this bug. However, the entire process took over two days.

Generating PoCs for an AI review poses two challenges. \ding{182}~\textit{Text review alone is insufficient to provide the context required to trigger the bug.} A review may identify suspicious code and explain why it could fail, but a PoC must construct the precise system-call sequence, kernel-object state, memory layout, and timing conditions that cause the bug to occur. Inferring this procedural trigger from the target patch alone requires substantially more exploration than recognizing suspicious code~\cite{anthropic2026exploit}.
\ding{183}~\textit{The trigger sequence and its execution environment are co-dependent requirements.} A valid PoC fails if the kernel configuration or userspace setup is wrong, yet these runtime requirements are rarely known in advance.
For example, file-system bugs may require specially formatted disk images and mount options~\cite{jbd2_corrupted_image}, while networking bugs may require virtual interfaces, namespaces, and protocol modules~\cite{ipvlan_skb_cb}.
Resolving them requires iterative execution. However, managing kernel builds, virtual machines, and error feedback simultaneously forces agents into unstructured retries, quickly exhausting their budget.

To this end, we present \sysname\footnote{We make \sysname public at \cite{artifact}.} (\textbf{K}ernel \textbf{A}I Review \textbf{V}ia \textbf{E}xecutable \textbf{R}eproduction), a multi-agent framework that closes the Review-to-PoC gap.

To address the first challenge, \sysname induces reusable reproduction strategies from successfully replayed historical bugs. These strategies capture how to construct a trigger, realize it as an executable PoC, and recognize whether the resulting fault matches the review.
To address the second challenge, we model PoC generation as a
search over a finite-state machine~(FSM). We observe that reproductions in the same subsystem go through similar steps, order, and failed places. Specialized procedures for different subsystems, such as loading BPF programs, can often be modeled as distinct FSM transitions.

\sysname's workflow is composed of two phases: an offline phase that learns the reproduction FSM for each subsystem from historical replays, and an online phase that searches over the strategies to generate a PoC for a new review.
\sysname uses two agents to explore the search space: a PoC agent that focuses on constructing the PoC, and an Environment agent that adapts the execution environment. The learned FSM governs the interaction between the two agents, ensuring that they explore the search space in a structured manner.
Each agent receives only the requirements of its
current state, which keeps its context on the step it is working on.
For the motivating case in \Cref{fig:intro_example}, \sysname generated a PoC in about one hour, while the original mailing-list thread took about 50 hours to progress from the textual review to maintainer acknowledgment.

We evaluated \sysname on 100 recent AI reviews from Sashiko and 100 known kernel bugs. In recent AI reviews, \sysname turns 49 AI reviews into PoC-backed reports, compared to 10 for an agentic reproducer and 9 for directed fuzzing. When these PoCs are submitted to kernel mailing lists,
\sysname increases the confirmation rate of Sashiko review from 9.3\% to 69.4\%. On the known bug benchmarks, \sysname generates PoCs with 89.5/100\% precision and 68/60\% PoC yield across the two benchmarks.
Ablation studies demonstrate that both reproduction FSMs and multi-agent environment adaptation framework play a critical role in PoC generation.

This paper makes the following contributions:
\begin{itemize}[leftmargin=*, labelindent=0pt]
    \item \textbf{General PoC Generation for AI kernel review.}
    We present a multi-agent framework that treats PoC generation and environment setup as a joint search problem. It continuously refines PoC alongside kernel configurations, emulated devices, and runtime resources, enabling AI reviews to support their findings with reproducible PoCs.
    \item \textbf{Reusable Kernel Reproduction Strategies.} We design a multi-agent process to replay historical kernel bugs and extract execution trajectories into reusable strategies, which capture trigger primitives, state transitions, and PoC realization steps for different kernel subsystems.
    \item \textbf{A large-scale evaluation with real-world impact.}
    \sysname outperforms agentic kernel bug reproducer and directed fuzzing tools on both recent AI-reviews~\textit{(49\% vs. 10\% vs 9\%)} and known ground-truth bugs~\textit{(64\% vs. 44\% vs. 29\%)}. Patches with \sysname-generated PoCs have received more positive responses~\textit{(71.4\% vs. 17.5\% vs. 19.5\%)} and faster fixes in the community.
\end{itemize}

\section{Background}
\begin{figure*}[t]
    \centering
    \includegraphics[width=1\textwidth]{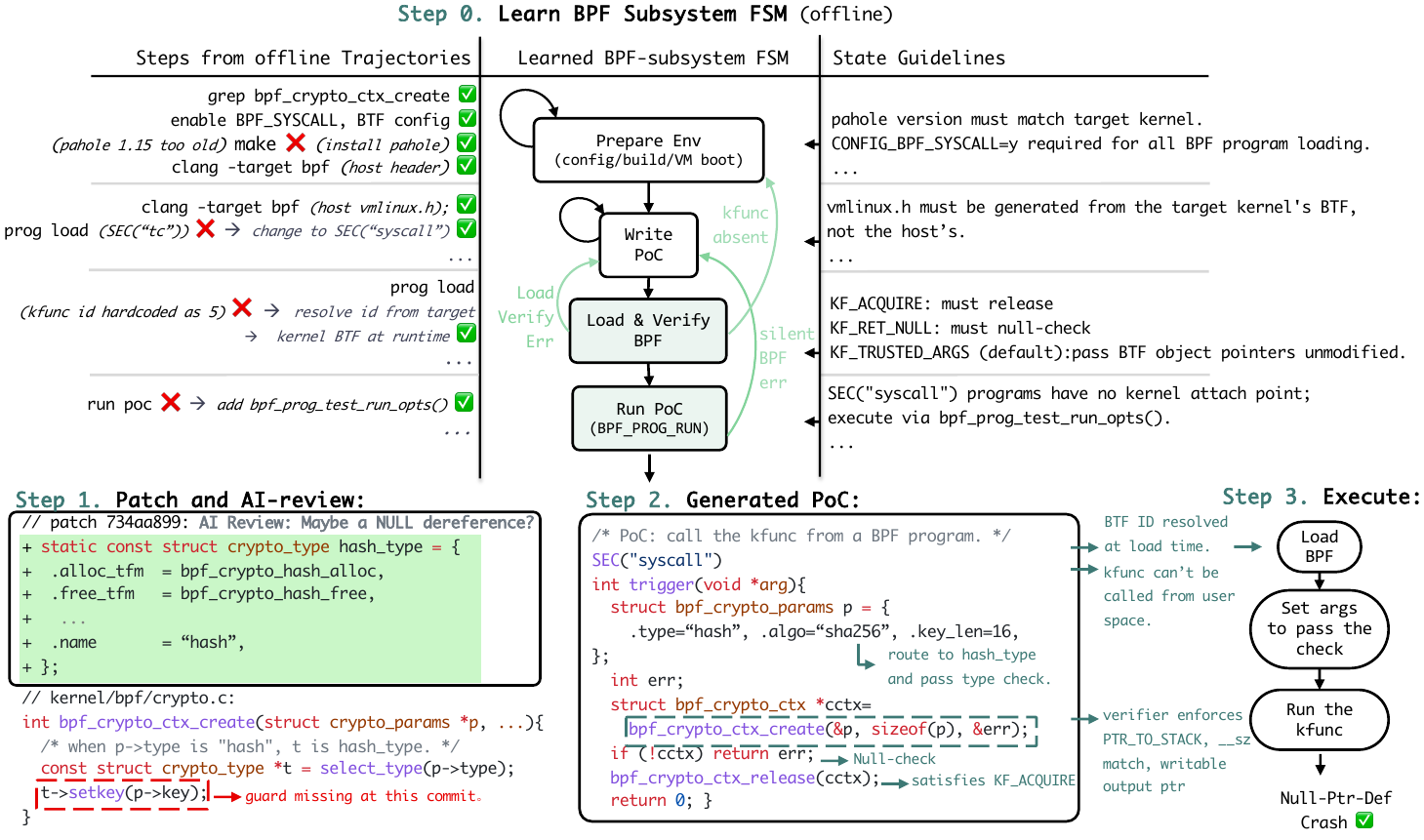}
    \caption{A motivating example: reproducing a NULL pointer dereference in
the BPF subsystem. Step 0 shows the learned BPF FSM induced offline from
historical trajectories---an FSM with attached. Steps 1--3 use this FSM online: given a patch and its AI review~(1), \sysname generates a PoC
by following the FSM~(2), and executes it to trigger the
reported crash~(3).
}
    \label{fig:motivating-bpf}
\end{figure*}
\textbf{Linux Kernel Maintenance Workflow.}
The Linux kernel is a massive distributed system maintained through a hierarchical, email-based patch workflow~\cite{kernel_process_doc}. Developers or maintainers propose code changes in the form of patches by emailing subsystem mailing lists (e.g., \textit{netdev}~\cite{netdev}, \textit{storage}~\cite{storage}). Subsystem maintainers review these patches, run sanity checks, and consolidate accepted changes into subsystem trees. Finally, Linus Torvalds pulls from these subsystem maintainers during specific merge windows to build the mainline kernel. With thousands of patches submitted weekly across hundreds of specialized subsystems, human review capacity is the primary bottleneck in kernel~\cite{lin2026learning}.

\textbf{Executable PoCs for AI Review.}
To ease this maintainer workload, LLM-based review agents~(e.g., \textit{Sashiko}) have recently been deployed at scale. These agents inspect submitted patches for logic errors, memory safety issues, and locking discipline, posting inline review comments to mailing lists. However, these agents act merely as speculative auditors: they output text comments asserting potential bugs without generating runnable proofs~\cite{sashiko,bpf-agent}. Converting a textual alert into verified evidence remains a human-dependent bottleneck, leaving maintainers overwhelmed by unverified alerts~\cite{ramachandra2025rfc}.

Confirming a reported bug requires constructing a proof-of-concept~(PoC)—a program or system-call sequence that drives the target kernel to reproduce the exact faulting call trace. Reproduction is inherently challenging due to:
\begin{enumerate}
\item \textit{Trigger Construction}: Triggering different bugs requires different system-call sequences, kernel-state transitions, and trigger primitives, such as allocation--free--reuse timing for a use-after-free or controlled thread interleavings for a race.

\item \textit{Environment Dependence}: Whether the vulnerable path is reachable may depend on the architecture, kernel configuration, boot parameters, emulated devices, resources, and userspace tools. These requirements are often discovered only through execution feedback.
\end{enumerate}

Existing reproduction methods span manual analysis, directed fuzzing, and exploit generation. Manual reproduction requires domain experts to painstakingly infer inputs and environments from scratch, making it labor-intensive and hard to scale. Automated kernel fuzzers~(e.g., \textit{Syzkaller~\cite{syzkaller2024}}) and directed greybox fuzzers~(e.g., \textit{SyzDirect~\cite{tan2023syzdirect}}) automate input generation toward target sites, but they explore state spaces without understanding free-form review comments and remain bounded by strict syscall grammars~(\textit{Syzlang}~\cite{syzkaller_syntax_doc}). Exploit generation tools~(e.g., \textit{FUZE}~\cite{FUZE}, \textit{SyzScope}~\cite{zou2022syzscope}) focus on escalating crash severity, but strictly assume a working crash input already exists.
Therefore, generating PoCs for a flagged review requires both understanding the suspicious bug and environment adaptation---the two capabilities combined in \sysname.

\section{Motivating Example}

\Cref{fig:motivating-bpf} shows a motivating example. The patch~(step 1) registers a new \codeztext{hash} crypto type, implementing the callbacks that hash algorithms need. However, it does not implement the \codeztext{setkey} type, since hash operations take no key, so that field is implicitly set to NULL.
The review says this can cause a Null-Ptr-Dereference by introducing this new type for the unimplemented \codeztext{setkey} type.
The problem lies in the function \codeztext{bpf\_crypto\_ctx\_create} that was written when \codeztext{hash} was not considered a registered type, and therefore \codeztext{setkey} will be called unconditionally.
It is hard for the maintainer to identify every potential crash line just from the patch files and review. And if the maintainer needs to verify the review by getting a PoC program, the review also does not say whether that path is reachable from user space, what input reaches it, or what kernel configuration is required. Faced with a review mail that can be neither confirmed nor dismissed, a maintainer must either spend hours investigating or set it aside.

\textbf{Two Coupled Requirements.}
Producing a PoC for this review requires two things.
The first is a suitable kernel build. Kernel config \codeztext{CONFIG\_DEBUG\_INFO\_BTF} must be enabled for loading the kfunc in BPF module. The version of \codeztext{pahole}, the BTF metadata generator, used during the build must also be compatible with the target kernel version.
The second is a sequence of valid BPF operations. To exploit a kfunc, not a syscall, the PoC must load a BPF program instead of calling syscall function from user space.
It must reference the kfunc using a dynamic BTF ID assigned by the target kernel, while satisfying unwritten verifier requirements—such as null-checking the context pointer before use and freeing it before exit.
These two conditions depend on each other. Without the correct kernel build, the BTF ID cannot be resolved, so the code fails to compile. Without valid BPF operations, the verifier rejects the program, so the kernel code is never executed. Fixing one without the other yields no progress.

\textbf{Why Existing Methods Fall Short.}
Kernel fuzzers (e.g., Syzkaller) fail to reach this path, not only due to limited testing efficiency, but also because of fundamental limitations in input representation. Syzkaller generates system calls defined in Syzlang, whereas calling a kfunc requires BPF bytecode. Although Syzkaller provides a kfunc-call instruction, Syzlang approximates its operand as \codeztext{int32[1:5]} because it cannot represent indices into the kernel's BTF table~\cite{syzkaller_bpf_specs}. Since the actual kfunc ID is dynamically assigned by the kernel and falls far outside this range, it's hard for fuzzing to generate the correct call. Consequently, neither Syzkaller nor SyzDirect triggered this path.

A general-purpose coding agent (e.g., Claude Code) must search both spaces at once without knowing the requirements for either. Because the patch introduces a new \codeztext{hash} type, the agent lacks prior context and must rely on iterative search. Every attempt is costly: changing a kernel config requires a full rebuild, while modifying the BPF program requires another load and verification cycle. Without clear direction, the agent fluctuates between configurations. It hesitates between Syzlang and C formats, revises the kernel config when loading fails, rebuilds, encounters verifier rejections, and retries. Ultimately, it eventually gets stuck in a repetitive loop.

\textbf{Generating the PoC with \sysname.}
\sysname guides the process using FSM learned from earlier BPF bug reproductions. \Cref{fig:motivating-bpf} (step 0) shows the BPF subsystem FSM extracted during the offline learning phase, which records the reproduction steps and their requirements. When the agents reach a state
, they receive its specific requirements directly. As a result, the kernel build is properly configured in a single round, and the agents know upfront that the target is a kfunc that must be invoked within a BPF program.

What remains is specific to this bug. As shown in \Cref{fig:motivating-bpf} (step 2), the agent reads the patch and surrounding code to determine which parameter values trigger the unguarded callback: setting \codeztext{.type} selects the new \codeztext{hash} algorithm, \codeztext{.algo} satisfies the validation check, and a non-zero \codeztext{.key\_len} ensures the code path reaches \codeztext{setkey}. With these fields populated, the program in \Cref{fig:motivating-bpf} (step 3) passes verification, executes, and triggers the target bug.

\section{Methodology}
\begin{figure*}[t]
    \centering
    \includegraphics[width=0.7\textwidth]{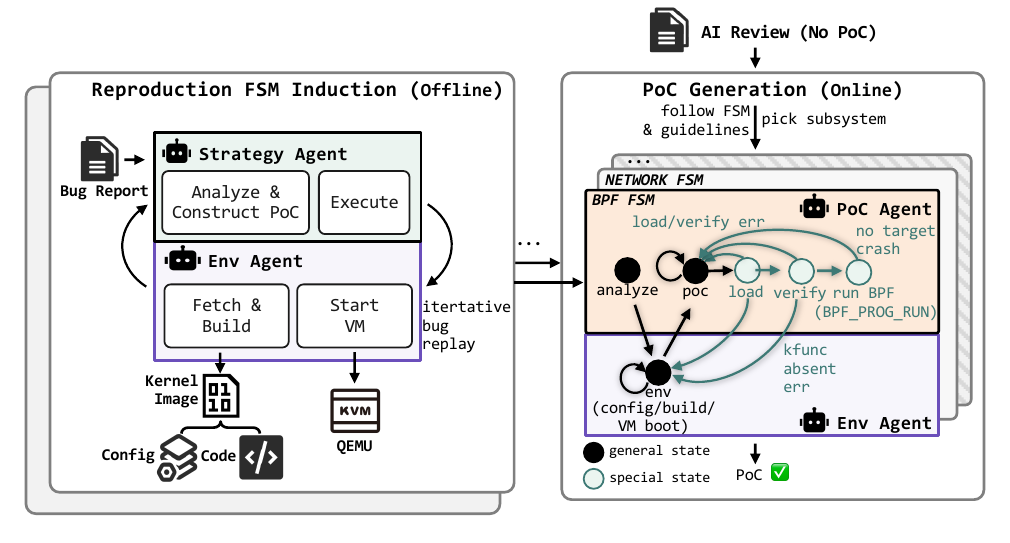}
    \caption{Overview of \sysname. In the offline phase~(left), the Strategy and Environment Agents replay historical bugs, and \Cref{alg:induction} derives one FSM per subsystem from successful trajectories. In the online phase~(right), \sysname selects the reviewed subsystem's FSM. Guided by this FSM and execution feedback, the PoC and Environment Agents iteratively refine the PoC and environment to produce an executable PoC for AI review.}
\label{fig:overview}
\end{figure*}

This section presents how \sysname generates PoCs for AI kernel reviews.
\Cref{sec:overview} gives an overview of the two-phase framework.
\Cref{sec:distill} describes how \sysname induces a reproduction FSM for each subsystem from the successful trajectories. \Cref{sec:dynamic_harness} describes how the PoC Agent and the Environment Agent use these FSMs to generate a PoC for a new AI review.

\label{sec:overview}

\Cref{fig:overview} shows the overall workflow of \sysname. The input is a textual AI review that identifies a potential bug without a PoC. \sysname works in two phases.
In the \emph{offline} phase, the Strategy Agent and the Environment Agent replay historical kernel bugs. They process successful trajectories and produce one reproduction FSM per subsystem.
In the \emph{online} phase, \sysname identifies the target patch subsystem and loads the corresponding FSM. The PoC Agent and the Environment Agent then follow the FSM, adjusting the PoC and the environment based on execution feedback until they trigger the reported bug.

\subsection{Reproduction Strategy Induction (Offline)}
\label{sec:distill}

A bug report describes what went wrong and where, but not how to trigger it. Reproducing a bug requires the right PoC program, the right kernel configuration, and the right VM setup, and some of these settings only become clear through execution. Inspired by how human experts reproduce kernel bugs, \sysname therefore replays historical bugs and records the full trace, not just the final result.

Across these replays, we observe two things. First, reproductions in the same subsystem follow a similar sequence of steps. The process starts with configuring and building the kernel, followed by booting the VM, writing and running a PoC, and checking its output.
For different subsystems, additional steps may be required. For example, reproducing a BPF bug may require loading and verifying a BPF program to invoke the relevant kfuncs, rather than using the usual system calls.
Second, failures during bug reproduction tend to occur at similar points, and each failure type returns the process to a specific earlier step~\cite{bouzenia2024repairagent, li2026syzharness}. Rejection of a verifier returns the process to the PoC step, whereas missing kernel option returns it to the environment build step. This structured order of steps and failure return points can be represented as a finite-state machine~(FSM).

\sysname therefore represents what it learns from one subsystem as an FSM whose states are the reproduction steps, whose edges are the returns after failure, and a set of guidelines attached to each state that record the failures observed and how they were resolved. \Cref{alg:induction} shows how to learn an FSM from reproduction trajectories.

\begin{table}[t]
\centering
\caption{Operation type rules. The four general types are fixed and shared
across subsystems.}
\label{table:optype}
\small
\begin{tabular}{@{}>{\raggedright\arraybackslash}p{\dimexpr0.72\columnwidth-\tabcolsep\relax}>{\raggedright\arraybackslash}p{\dimexpr0.28\columnwidth-\tabcolsep\relax}@{}}
\toprule
\textbf{Operation} & \textbf{Type} \\
\midrule
Read kernel source, reviews, patches                & \texttt{analyze} \\
\arrayrulecolor{gray!20}\midrule\arrayrulecolor{black}
Change kernel config, build kernel,     & \texttt{env} \\
install tools, boot the VM                    & \\
\arrayrulecolor{gray!20}\midrule\arrayrulecolor{black}
Write, compile or adjust the PoC                      & \texttt{poc} \\
\arrayrulecolor{gray!20}\midrule\arrayrulecolor{black}
Execute the PoC, read the VM log              & \texttt{run} \\
\arrayrulecolor{gray!80}\midrule\arrayrulecolor{black}
Anything else                                 & \texttt{special:cmd} \\
\bottomrule
\multicolumn{2}{@{}p{0.98\columnwidth}@{}}{\footnotesize
For example, \codeztext{bpftool prog load} in BPF trajectories and
\codeztext{mkfs} or \codeztext{mount} in file-system trajectories match no
general rule and are therefore named after the command itself.
} \\
\end{tabular}
\end{table}

\begin{algorithm}[t]
\caption{Reproduction Subsystem FSM Induction}
\label{alg:induction}
\small
\begin{algorithmic}[1]
\Require Successful replay trajectories $T$ of one subsystem
\Ensure FSM $(N, E)$ and its state guidelines $G$

\ForAll{step $x$ in each $\tau \in T$}
    \State $\ell(x) \gets$ type from \Cref{table:optype},
           or the command name if no rule matches
\EndFor
\State $N \gets \{\, \ell : \ell \text{ occurs in} \geq 2
       \text{ trajectories of } T \,\}$
\State $E \gets \{
    \texttt{analyze} \to \texttt{env},
    \texttt{env} \to \texttt{poc},
    \texttt{poc} \to \texttt{run}
    \}$
\State $G \gets \emptyset$

\ForAll{step that fails at state $A \in N$}
    \If{$A$ is later reached again and succeeds}
        \State $R \gets$ set of all states between the failure and success
        \ForAll{$r \in R$}
            \State $E \gets E \cup \{\, A \to r \,\}$
            \State $G \gets G \cup
                   \{\, r \mapsto (\text{error},\ \text{fix}) \,\}$
        \EndFor
    \EndIf
\EndFor
\State remove entries in $E$ and $G$ with fewer than 2 occurrences
\end{algorithmic}
\smallskip
{\footnotesize
\noindent\textit{Note:}
$(\mathrm{err}, \mathrm{fix})$ denotes the error observed at state $A$
and the corrective action taken at state $r$, respectively.
\par}
\end{algorithm}

\textbf{Historical Bug Replay.}
Given a historical bug report and its corresponding kernel version, as shown on the left side of \Cref{fig:overview}, the Strategy Agent and the Environment Agent attempt to reproduce the given bug. The Strategy Agent analyzes the report, constructs the PoC, and executes it. The Environment Agent builds the kernel, boots the VM, and adjusts the configuration when execution requests it. The agents iterate until the PoC triggers the expected fault. The PoC program can be written
in C or Syzlang~\cite{syzkaller_syntax_doc} format. Syzlang covers most system calls directly, while C handles cases that require arbitrary control flow or subsystem-specific interfaces, such as BPF programs~\cite{syzkaller_pseudo_syscalls}.

\textbf{FSM Output.}
We collect 175 Syzbot~\cite{syzbot} reports and 631 historical kernel CVEs as candidates, covering the subsystem distribution in the kernel.
Syzbot bugs are automatically discovered by kernel fuzzing, whereas CVE bugs are identified as security-relevant. We remove duplicates between the two sources. These historical bugs represent high-priority bugs identified by both automated tools and manual analysis.
Only cases where the replay reaches the expected fault contribute to the output.
Successfully reproduced bugs are stored as trajectories, each of which records every tool call made by the agent and its corresponding result.

\begin{table*}[htbp]
\centering
\caption{Historical kernel bugs collected for reproduction FSMs induction.
}
\label{table:distill_data}
\small
\begin{tabular}{
>{\raggedleft\arraybackslash}m{0.2\columnwidth}
m{0.25\columnwidth}
>{\arraybackslash}m{0.2\columnwidth}
m{0.9\columnwidth}}
\toprule
\makecell[r]{\textbf{Kernel}\\\textbf{Subsystem}} &
\makecell[r]{\textbf{Historical Bugs}\\\textit{(Syzbot+CVE)}} &
\makecell[l]{\textbf{Successful}\\\textbf{Trajectories}} &
\makecell[l]{\textbf{Example}}
\\
\midrule
Core Kernel          & 62~(\textit{11+51}) & 39       & Race syscall lifecycle paths (\codeztext{fork/exec/exit}, \codeztext{fd teardown}) to reach lock and RCU code. \\
\arrayrulecolor{gray!20}\midrule\arrayrulecolor{black}
\makecell[r]{Memory\\Management}    & 203~(\textit{7+196}) &  9      & Drive fault and allocation edge cases through \codeztext{mmap}/\codeztext{madvise}/\codeztext{truncate} with hugepages configured. \\
\arrayrulecolor{gray!20}\midrule\arrayrulecolor{black}
File System          & 162~(\textit{78+84})& 97      & Craft a disk image with the target feature layout, then mount it with matching options. \\
\arrayrulecolor{gray!20}\midrule\arrayrulecolor{black}
Network              & 226~(\textit{46+180})& 112     & Set up virtual NICs (mac80211\_hwsim) and drive protocol state machines through netlink. \\
\arrayrulecolor{gray!20}\midrule\arrayrulecolor{black}
BPF                  & 12~(\textit{5+7})&  10          & Load a BPF program whose kfunc call is resolved from the target BTF, then run \codeztext{BPF\_PROG\_TEST\_RUN}. \\
\arrayrulecolor{gray!20}\midrule\arrayrulecolor{black}
Drivers              & 32~(\textit{21+11})& 24       & Attach a QEMU virtual device and force hardware features through module parameters. \\
\arrayrulecolor{gray!20}\midrule\arrayrulecolor{black}
Firmware             & 10~(\textit{1+9})&   6         & Feed malformed device-tree data through QEMU's -dtb option to reach property validation. \\
\arrayrulecolor{gray!20}\midrule\arrayrulecolor{black}
Other                & 99~(\textit{6+93}) & 20 & Emulate the target architecture (\codeztext{powerpc}/\codeztext{riscv}/\codeztext{s390}) under QEMU TCG. \\
\arrayrulecolor{gray!80}\midrule\arrayrulecolor{black}
\multicolumn{1}{r}{Total} & 806~(\textit{175+631}) & 317 & \\
\bottomrule
\end{tabular}
\end{table*}

\Cref{alg:induction} describes how each successful trajectory is used to produce a subsystem FSM$(N, E)$. State guidelines in each node include important details and lessons from trial and error.
The key observation behind the algorithm is that most PoC reproduction processes share a small set of high-level steps regardless of the target subsystem: reading the code to understand the bug~(\codeztext{analyze}), setting up the kernel and VM~(\codeztext{env}), writing and running the PoC, and checking the output~(\codeztext{poc} and \codeztext{run}).
\Cref{table:optype} formalizes these four general operation types. Any step during the reproduction that does not fit a general type is named after itself~(Line 2). Then a set of similar steps becomes a subsystem-specific state---for example, loading and verifying a BPF program in BPF trajectories or \codeztext{mount} in file-system subsystems.

In the general FSM composed of the four basic states above, if a special state~(A in Line 7) associated with a step fails initially but succeeds after the agent fixes the issue and retries the step, edges from A to all states between A and the success state are added~(Line 7-11).
When a subsystem frequently repeats a specialized type workflow, we elevate that procedure into a distinct node and its edges within its FSM, and the new specialized FSM is generated.

Even at the same FSM node, the exact steps executed often vary between different subsystems. We therefore analyze historical execution steps to derive targeted state key requirements, including kernel configurations, virtual machine setups, PoC attempt budgets, and so on. Whenever the agent enters a specific node, the corresponding knowledge is dynamically injected into its prompt context.

\Cref{table:distill_data} shows the learned subsystem FSMs from the offline bug replay. We summarize a customized FSM for each subsystem to capture its unique bug reproduction steps.
Each subsystem also incorporates state guidelines that cover the specifics of setting up the environment and writing the PoC.

For bugs that are difficult to categorize, we apply a general FSM with states \codeztext{analyze}, \codeztext{env}, \codeztext{poc} and \codeztext{run}, as shown in \Cref{table:optype} while still summarizing dedicated guidelines for them. Each state includes subsystem-specific requirements across different configurations and architectures for that state.
Network and Drivers often need to handle different types of virtual devices, while File System involves different file-system implementations with varying mount, initialization, and configuration requirements.

\Cref{fig:motivating-bpf} shows the simplified FSM example induced for the BPF subsystem. The black general states (\codeztext{analyze}, \codeztext{env}, \codeztext{poc}, \codeztext{run}) appear in every subsystem. The green states~(\codeztext{load}, \codeztext{verify}) are specific to BPF, because a BPF program needs to be loaded and verified before execution, and these phases are learned and distilled into the FSM. The green edges come from observed failures: a verifier rejection at \codeztext{load} sends the agent back to \codeztext{poc} to revise the program, while a missing BTF configuration at \codeztext{load} sends it back to \codeztext{env} to reconfigure the kernel.

The guidelines on the right side of the table list the specific error information and fixes recorded from those failures of each state. For instance, the version of the BTF metadata generator, \codeztext{pahole}, must match the target kernel version. The BPF information file \codeztext{vmlinux.h} must be generated from the target kernel's BTF, and the kernel reference acquisition flag \codeztext{KF\_ACQUIRE} references must be released before the program exits.
These important details are extracted and stored in FSMs, which provide essential guidance for kernel building, VM configuration, and PoC development and execution.

\subsection{Kernel PoC Generation (Online)}
\label{sec:dynamic_harness}

When given a new AI review that reports a potential kernel bug, \sysname identifies the subsystem of the reported bug and loads the corresponding FSM. The PoC Agent and the Environment Agent then follow the FSM to generate a PoC and trigger the bug under a suitable environment.

\textbf{Following the FSM.}
The FSM controls which agent acts at each step and how failures are handled. The FSM does not strictly constrain every decision made by the agent. It provides a framework that separates LLM reasoning from the required steps of PoC generation, allowing the LLM to search and analyze more efficiently.
The system includes two agents: the PoC Agent and the Environment Agent.
The PoC Agent handles the states \codeztext{analyze}, \codeztext{poc}, and \codeztext{run}, which are related to PoC exploration; the Environment Agent handles the state \codeztext{env} for environment setup searching.
For subsystem-specific states, those related to kernel building, QEMU configuration, and booting are assigned to the Environment Agent, while the remaining states are assigned to the PoC Agent.
For example, special states \codeztext{load} and \codeztext{verify} in the BPF FSM shown in \Cref{fig:motivating-bpf} are assigned to PoC Agent for the BPF program loading and verification.

The FSM also enforces the order of some state transitions. The kernel must be built after analyzing the review mail and kernel source code, and the VM must be ready before the PoC can run. When execution fails, the back edges determine the next state, whether to rebuild the kernel or revise the current PoC code. For example, a verifier rejection during \codeztext{load} sends the PoC Agent back to \codeztext{poc} to revise the program. A missing kernel configuration sends the Environment Agent back to state \codeztext{env} to reconfigure and rebuild the kernel. During this process, the PoC Agent waits for the Environment Agent to finish and resumes when the updated environment is ready. This design separates code exploration from complex environment configuration, preventing the agent from being overwhelmed by interleaved environment logs and code.

\begin{table*}[htbp]
\centering
\caption{Taxonomy of tools available to the Environment Agent and PoC Agent.}
\label{table:tools}
\small
\begin{tabular}{@{}m{0.21\columnwidth}>{\raggedright\arraybackslash}m{0.4\columnwidth}>{\raggedright\arraybackslash}m{0.4\columnwidth}m{0.7\columnwidth}@{}}
  \toprule
  \textbf{Category} & \textbf{Environment Agent Tools} & \textbf{PoC Agent Tools} & \textbf{Description} \\
  \midrule
  Environment &
  \code{fetch}, \code{config}, \code{build\_kernel}, \code{start/stop\_qemu} &
  $\varnothing$ &
  Env side manages the full lifecycle from  source acquisition to QEMU boot; PoC side triggers rebuild/restart via cross-agent signaling. \\
  \arrayrulecolor{gray!20}\midrule\arrayrulecolor{black}
  File Operations &
  \code{find\{n\}}, \code{grep\{n\}}, \code{list\{n\}}, \code{read\{n\}}, \code{write\_new} &
  \code{find\{n\}}, \code{grep\{n\}}, \code{list\{n\}}, \code{read\{n\}}, \code{write\_new} &
  Shared file-system primitives providing search, navigation, read--write access. \\
  \arrayrulecolor{gray!20}\midrule\arrayrulecolor{black}
  PoC Execution &
  $\varnothing$ &
  \code{compile\_poc}, \code{copy\_to\_vm}, \code{execute\_poc}, \code{get\_vm\_log} &
  Compiles PoC source into executable artifacts, executes it in QEMU and collects output. \\
  \arrayrulecolor{gray!20}\midrule\arrayrulecolor{black}
  Flow Control &
  \code{summary} &
  \code{analyze}, \code{stop}, \code{finish}, \code{summary} &
  Drives phase progression and termination of the agent workflow and triage bugs. \\
  \bottomrule
  \end{tabular}
  \end{table*}

For each state in FSMs, guidelines serve as the prompts that are injected into the agent's context. Using the BPF FSM in
\Cref{fig:motivating-bpf} as an example: when the Environment Agent
enters the state \codeztext{env}, it receives rules such as ``\textit{pahole version must match the target kernel}'' and ``\textit{CONFIG\_DEBUG\_INFO\_BTF requires dwarves
greater than 1.16}''. When the PoC Agent enters poc, it receives ``\textit{compile with clang -target bpf, not {gcc}}'' and ``\textit{vmlinux.h must come from the target kernel's BTF}''.
These rules and knowledge give the agent a concrete starting point and help it avoid failures that were already observed during offline replay.

\subsection{Implementation}

We implement \sysname in Python. Each subsystem FSM is stored in a JSON file, which records the FSM states, valid transitions, and the guidelines attached to that state. At the start of a run, \sysname identifies the subsystem of the target patch by examining the patch's metadata and loads the corresponding FSM.

The FSM controller invokes one agent at a time.
It also exposes only the tools available in that state. The Environment Agent handles kernel configuration, compilation, tool installation, and QEMU management. The PoC Agent handles source analysis, PoC construction, compilation, deployment, execution, and log collection. File operations are available to both agents, as shown in \Cref{table:tools}.

When the PoC requires an environment change, the controller pauses the PoC Agent and invokes the Environment Agent. After completing the build or VM update, the Environment Agent returns a short summary of the new environment. The controller replaces the previous environment information and removes execution results that no longer apply. The PoC Agent then continues from the state selected by the FSM. A run ends when the PoC triggers the reported bug or reaches its execution budget.

\section{Evaluation}
We aim to address the following research questions:
\begin{itemize}[leftmargin=*, labelindent=0pt]
    \item \textbf{RQ1~(Real-world Impact):}
    How effectively does \sysname generate executable PoCs for recent AI kernel reviews, and does it assist Linux maintainers?
    \item \textbf{RQ2~(Verification Accuracy):} How effectively and precisely does \sysname generate PoCs for kernel bugs?
    \item \textbf{RQ3~(Ablation and Mechanism Analysis):} How do reproduction FSM and the two-agent framework contribute to \sysname's effectiveness and efficiency?
\end{itemize}

\subsection{Experimental Setup}
\subsubsection{Datasets}
For RQ1 and RQ3, we collect the 100 most recent review--patch pairs flagged as problematic by Sashiko~\cite{sashiko}, a kernel agentic review system maintained by Google. Using recent reviews evaluates PoC generation on current kernel code and reduces the risk that an LLM has memorized widely discussed historical vulnerabilities. For RQ2, we collect 50 Syzbot reports~\cite{syzbot} and 50 historical kernel CVEs with confirmed triggering scenarios.
These 100 evaluation bugs are disjoint from the historical bugs used for offline replay. None of these evaluation bugs or their reproduction trajectories contributed to the construction of the subsystem FSMs or state guidelines.
All evaluated methods receive only the target kernel revision and the corresponding AI review or bug report; existing target-specific PoCs and reproduction trajectories are withheld.

\subsubsection{Baselines and Evaluation Protocol}
We compare \sysname with K-Repro~\cite{krepro}, an agentic kernel bug reproducer, and SyzDirect~\cite{tan2023syzdirect}, a directed greybox kernel fuzzer. They represent emerging AI methods for bug reproduction and traditional testing methods, respectively.
RQ3 compares against a general-purpose coding agent, Claude Code, and ablated variants of \sysname. SyzDirect runs for 24 hours per case. For LLM-based methods, each task is limited to a maximum of 500 interaction rounds and 100 million tokens. We record a \mbox{@Crash} when a generated PoC triggers a kernel fault under KASAN~\cite{KernelAddressSanitizer}, and a \mbox{@Match} when the resulting fault and call trace conform to the failure described by the review or ground-truth bug report.

\subsubsection{Implementation Environment}
We conducted the experiments on Ubuntu 22.04 LTS with an AMD EPYC 9534 64-core processor and 256\,GB RAM. We use GCC 12.4.0 for kernel build, QEMU 8.2.2 as the virtualization platform, and DeepSeek-V4-Flash~(April 24, 2026) and Qwen-3.8-Flash as the base model for all LLM-based methods. While the host platform remains fixed, \sysname can adapt the guest architecture, kernel configuration, emulated devices, CPU and memory resources, and tools for each target.

\subsection{RQ1: Real-world Impact of \sysname}
\begin{table*}[t]
\centering
\caption{Verification Results on 100 Sashiko AI Reviews.}
\label{table:rq1a}
\small
\setlength{\tabcolsep}{7pt}
\begin{tabular}{@{}
  >{\raggedright\arraybackslash}m{0.145\textwidth}
  *{2}{
    >{\raggedleft\arraybackslash}m{\dimexpr0.14\textwidth-2\tabcolsep\relax}
    >{\raggedright\arraybackslash}m{\dimexpr0.14\textwidth-2\tabcolsep\relax}
  }
  >{\centering\arraybackslash}m{\dimexpr0.15\textwidth-2\tabcolsep\relax}
@{}}
\toprule
\makebox[\linewidth][c]{\multirow{2}{*}[-0.7ex]{\textbf{Metric}}} & \multicolumn{2}{c}{\textbf{\sysname}} & \multicolumn{2}{c}{\textbf{K-Repro}\cite{krepro}} & \multirow{2}{*}[-0.7ex]{\textbf{SyzDirect}\cite{tan2023syzdirect}} \\
\arrayrulecolor{gray!20}\cmidrule(lr){2-3}\cmidrule(lr){4-5}\arrayrulecolor{black}
& {\fontsize{8.5}{10}\selectfont\mdseries\mbox{D4VS}} & {\fontsize{8.5}{10}\selectfont\mdseries\mbox{Qwen3.8}} & {\fontsize{8.5}{10}\selectfont\mdseries\mbox{D4VS}} & {\fontsize{8.5}{10}\selectfont\mdseries\mbox{Qwen3.8}} & \\
\midrule
\makebox[\linewidth][l]{\textbf{@Crash}\hfill\textbf{(\%)}} & \textbf{52.0}~(\textit{52/100}) & 45.0~(\textit{45/100}) & 13.0~(\textit{13/100}) & 11~(\textit{11/100}) & 27.0~(\textit{27/100}) \\
\makebox[\linewidth][l]{\textbf{@Match}\hfill\textbf{(\%)}} & \textbf{49.0}~(\textit{49/100}) & 42.0~(\textit{42/100}) & 10.0~(\textit{10/100}) & 10~(\textit{10/100}) & 9.0~(\textit{9/100}) \\
\arrayrulecolor{gray!20}\midrule\arrayrulecolor{black}

\makebox[\linewidth][l]{\textbf{Time}\hfill\textbf{(min)}} & \textbf{72.8} & 250.1 & 120.2 & 372.6 & 1440.0 \\
~~\textit{-- Matched} & 62.7 & 237.3 & 50.0 & 205.5 & 1440.0 \\
~~\textit{-- Unmatched} & 82.6 & 259.4 & 128.0 & 391.2 & 1440.0 \\
\arrayrulecolor{gray!20}\midrule\arrayrulecolor{black}

\makebox[\linewidth][l]{\textbf{Input Tokens}\hfill\textbf{(M)}} & \textbf{13.0} & 27.05 & 20.8 & 43.7 & - \\
~~\textit{-- Matched} & 8.9 & 20.20 & 10.0 & 15.2 & - \\
~~\textit{-- Unmatched} & 17.0 & 32.01 & 22.0 & 46.9 & - \\
\arrayrulecolor{gray!20}\midrule\arrayrulecolor{black}
\makebox[\linewidth][l]{\textbf{Output Tokens}\hfill\textbf{(K)}} & \textbf{274.8} & 830.0 & 339.0 & 593.9 & - \\
~~\textit{-- Matched} & 234.1 & 790.8 & 150.0 & 283.5 & - \\
~~\textit{-- Unmatched} & 313.9 & 858.4 & 360.0 & 628.4 & - \\
\arrayrulecolor{gray!20}\midrule\arrayrulecolor{black}
\makebox[\linewidth][l]{\textbf{Cost}\hfill\textbf{(\$)}} & \textbf{0.210} & 2.077 & 0.308 & 3.055 & - \\
~~\textit{-- Matched} & 0.156 & 1.620 & 0.144 & 1.093 & - \\
~~\textit{-- Unmatched} & 0.262 & 2.408 & 0.326 & 3.273 & - \\
\arrayrulecolor{gray!20}\midrule\arrayrulecolor{black}

\makebox[\linewidth][l]{\textbf{Agent Steps}} & \textbf{136.0} & 184.8 & 399.5 & 471.2 & - \\
~~\textit{-- Matched} & 111.0 & 173.6 & 170.0 & 213.3 & - \\
~~\textit{-- Unmatched} & 160.0 & 192.9 & 425.0 & 499.9 & - \\

\bottomrule
\multicolumn{6}{@{}p{0.9\textwidth}@{}}{
\textbf{Note:}
\textbf{D4VS} means DeepSeek-V4-Flash, \textbf{Qwen3.8} means Qwen-3.8-Flash.
\textit{@Crash} denotes that a kernel crash was successfully triggered, while
\textit{@Match} signifies that the triggered crash call trace aligns with the bug in the review.
}
\end{tabular}
\end{table*}

To evaluate whether \sysname can turn recent AI kernel reviews into executable PoC evidence and whether the resulting PoC-backed reports assist kernel maintainers. We evaluate \sysname on 100 recent Sashiko review--patch pairs and subsequently submit the generated PoCs to the corresponding mailing-list threads. The recency of these cases provides in-the-wild targets from current kernel development while reducing overlap with historical vulnerabilities.

To our knowledge, K-repro is the only agentic bug reproducer for the Linux kernel. Because K-Repro is not publicly available, we re-implement it using the same model as in \sysname, following its published design. We treat \mbox{@Match} as the primary metric. Triggering an arbitrary kernel crash does not mean that the PoC is correct unless the resulting fault matches the failure described in the review. To ensure experimental rigor, we manually re-check every matched crash log against the corresponding review.
The two authors independently assessed all the crash-producing cases.
\vspace{-1ex}
\subsubsection{Bug Finding on AI Reviews}
\Cref{table:rq1a} reports on the PoC generation results. \sysname triggers kernel crashes for 52 of the 100 reviews and produces matching PoCs for 49, meaning that 94.2\% of its triggered crashes support the corresponding review claims. With DeepSeek-V4-Flash, \sysname takes 72.8 minutes and costs \$0.210 per case on average. The 49 matched cases are completed in 62.7 minutes at \$0.156 per case, whereas the 51 unmatched cases average 82.6 minutes and \$0.262 per case. Overall averages are recomputed from the rounded subgroup averages using the corresponding case counts. The three mismatches occur when \sysname triggers similar faults in other code paths and incorrectly associates them with the reviewed defects.

SyzDirect triggers 27 crashes within its 24-hour budget per case, but demonstrates the reviewed failures in only 9 cases. This result reflects a mismatch between directed fuzzing and the needs of PoC-backed review. SyzDirect searches toward a target code location, whereas a review often describes a particular failure mechanism and requires specific kernel state, making complex subsystem-specific triggers, e.g., BPF, and userspace interactions difficult to realize. And it logs any kernel crash in its final results, so maintainers still need to re-check the results and dismiss the false positives.

Out of 100 AI-generated reviews, 61 demanded non-default environments~(\textit{x86}), spanning alternative architectures (\textit{\mbox{AArch64}/\allowbreak\mbox{RISC-V}/\allowbreak\mbox{ARM32}}), multi-core setups (for race conditions), large memory allocations (for retries), and specialized userspace utilities (e.g., \textit{BlueZ} for Bluetooth handling). Without the reproduction FSM and two-agent harness, K-Repro achieved 10 successful matches under the baseline configuration.
When the LLM encountered these environment mismatches, it fell into blind retry loops—repeatedly attempting kernel rebuilds and QEMU boots. This erratic behavior led to an inflated step count~(399.5 vs. 136.0) and a substantial escalation in LLM API costs~(0.308\$ vs. 0.210\$).

\begin{table}[htbp]
\centering
\caption{Results by Kernel Subsystem on Sashiko Reviews.}
\label{table:rq1b}
\small
\begin{tabular}{@{}
  >{\raggedleft\arraybackslash}m{\dimexpr0.245\columnwidth-\tabcolsep\relax}
  >{\centering\arraybackslash}m{\dimexpr0.18\columnwidth-2\tabcolsep\relax}
  >{\centering\arraybackslash}m{\dimexpr0.18\columnwidth-2\tabcolsep\relax}
  >{\centering\arraybackslash}m{\dimexpr0.205\columnwidth-2\tabcolsep\relax}
  >{\centering\arraybackslash}m{\dimexpr0.19\columnwidth-\tabcolsep\relax}
@{}}
\toprule
\textbf{Subsystem} & \textbf{Reviews} & \textbf{\sysname}  & \textbf{SyzDirect} & \textbf{K-Repro}\\
\midrule
Kernel            & 10 & 5  & 1 & 1 \\
\arrayrulecolor{gray!20}\midrule\arrayrulecolor{black}
Memory            & 10 & 6  & 1 & 1\\
\arrayrulecolor{gray!20}\midrule\arrayrulecolor{black}
File              & 24 & 12 & 4 & 4 \\
\arrayrulecolor{gray!20}\midrule\arrayrulecolor{black}
Network           & 23 & 11 & 1& 3  \\
\arrayrulecolor{gray!20}\midrule\arrayrulecolor{black}
BPF               & 13 & 6  & 0 & 1  \\
\arrayrulecolor{gray!20}\midrule\arrayrulecolor{black}
Drivers           & 10 & 5  & 0 & 0  \\
\arrayrulecolor{gray!20}\midrule\arrayrulecolor{black}
Firmware          & 10 & 4  & 2 & 0  \\
\arrayrulecolor{gray!80}\midrule\arrayrulecolor{black}
\textbf{Total}    & 100 & 49 & 9 & 10 \\
\bottomrule
\multicolumn{5}{@{}p{\columnwidth}@{}}{\footnotesize Entries report the number of reviews for which each method produces a PoC demonstrating the reported failure.}
\end{tabular}
\end{table}

\Cref{table:rq1b} breaks down the PoC generation results by kernel subsystem on DeepSeek-V4-Flash. \sysname produces PoCs that demonstrate the failures reviewed in all seven subsystems, covering 40--60\% of the reviews in each category and outperforming both baselines throughout. This consistent performance indicates that reproduction FSMs distilled from the offline phase and environment adaptation are applicable beyond a single class of kernel targets. Firmware remains the most challenging category, with PoCs generated for 4 of 10 reviews, reflecting the difficulty of reproducing hardware- and platform-dependent behavior in QEMU. Driver cases generally require specialized emulated devices and iterative adjustment of QEMU parameters; neither SyzDirect nor K-Repro successfully produces a PoC in this category. For the BPF subsystem, SyzDirect does not produce PoC because its Syzlang-based input model provides limited support to express the BPF operations required by these reviews~\cite{syzkaller_syntax_doc}.

\subsubsection{Maintainer Reception}
\vspace{-2ex}
\begin{table}[htbp]
\centering
\caption{Reply Rate to Kernel Review/Report Emails.}
\label{table:rq-replyrate}
\small
\begin{tabular}{
>{\raggedright\arraybackslash}m{0.2\columnwidth}
>{\centering\arraybackslash}m{0.2\columnwidth}
>{\centering\arraybackslash}m{0.28\columnwidth}
}
\toprule
\textbf{System}     & \textbf{Replies}  & \textbf{Confirmed Bugs}\\
\midrule
Sashiko & 21.2\% & 9.3\% \\
\arrayrulecolor{gray!20}\midrule\arrayrulecolor{black}
\sysname     & \textbf{71.4\%} & \textbf{69.4\%} \\

\arrayrulecolor{gray!20}\midrule\arrayrulecolor{black}
Syzbot & 19.5\% & 16.4\% \\
\bottomrule
\end{tabular}
\end{table}

Beyond aggregate metrics, the practical value of \sysname is measured by whether its generated PoCs assist kernel maintainers in their review workflow.
Of the 49 reviews where \sysname achieved a verified match, we submitted the corresponding PoCs to the kernel mail threads.
Maintainers replied to 35 of our correspondence emails, explicitly confirming that 34 of them correctly demonstrated the bugs described in the AI reviews and were fixed. The remaining one PoC was not accepted because it triggers a crash through brute-force attempts, making it of limited practical value.

We collected one month of bug reports sent by Sashiko and Syzbot to the Linux kernel community and measured the email reply and bug confirmation rates. To fairly compare \sysname with Sashiko and Syzbot in terms of maintainer reply willingness and bug confirmation, we sampled 100 emails per round evenly across the subsystems~(\Cref{table:rq1b}), repeated this 10 times, averaged the results, and reported them in \Cref{table:rq-replyrate}.
In comparison, Sashiko and Syzbot emails received response rates of 21.2\% and 19.5\%, with confirmation rates of 9.3\% and 16.4\%, respectively.
Although \sysname builds on Sashiko review by additionally sending verifiable PoCs, under sampling with the same base number, this still partially demonstrates that \sysname can make kernel textual reviews easier to assess and act upon.

The generated PoCs also revealed the practical status of several reviewed defects. In one of our generated PoCs, the patch containing a reported use-after-free had already been merged into the mainline kernel before \sysname produced executable evidence for it. In two other cases, the PoCs demonstrated that reported out-of-bounds and race defects remained reproducible in newer revisions despite earlier fixes. These cases illustrate how PoC generation for AI review helps maintainers determine whether a reviewed bug still affects current code and prioritize corrective action.

\begin{tcolorbox}[boxrule={0.03cm}, boxsep={0.01cm}]
\textbf{RQ1 Summary:} \sysname verified 49 of 100 recent AI reviews with call-trace-matching PoCs, compared with 10 for K-Repro and 9 for SyzDirect. Maintainers replied to 35 of the 49 submitted reports and confirmed 34, showing that PoC generation makes AI reviews more credible and actionable.
\end{tcolorbox}

\vspace{-1ex}
\subsection{RQ2: PoC Generation on Ground-Truth Bugs}
Although the latest AI reviews can test \sysname’s ability to validate new bugs, the problem with Sashiko review is that it lacks a true oracle. In other words, we do not know how many bugs were overlooked.

RQ2 evaluates \sysname on 50 Syzbot reports and 50 historical kernel CVEs with confirmed triggering scenarios. This ground-truth setting allows us to measure how often \sysname reconstructs a PoC for a known bug and how reliably the generated crashes reproduce the intended fault. Each method receives only the target kernel source code and bug report, while existing target-specific PoCs are withheld. And all 50 Syzbot reports and 50 CVEs are not included in the reproduction strategy induction process of \sysname.

\begin{table}[t]
  \centering
  \caption{Verification Results on Syzbot and CVEs.}
  \label{table:rq2a}
  \small
  \setlength{\tabcolsep}{4pt}

  \noindent\textbf{(a) Syzbot Benchmark (n=50)}
  \vspace{4pt}
  \begin{tabular}{@{}l c c c c c @{}}
  \toprule
  \textbf{Method} & \textbf{Verified} & \textbf{Mismatched} & \textbf{Missed} & \textbf{Prec.} & \textbf{Poc.} \\
\arrayrulecolor{gray!20}\midrule\arrayrulecolor{black}
  \multicolumn{6}{@{}l}{\sysname} \\
  \quad \textit{D4VS}      & 34 & 4 & 12 & 89.5\% & 68.0\% \\
  \quad \textit{Qwen3.8}   & 28 & 9 & 13 & 75.7\% & 56.0\% \\

\arrayrulecolor{gray!20}\midrule\arrayrulecolor{black}

  \multicolumn{6}{@{}l}{K-Repro} \\
  \quad \textit{D4VS }     & 23 & 15 & 12 & 60.5\% & 46.0\% \\
  \quad \textit{Qwen3.8} & 26 & 12 & 12 & 68.4\% & 52.0\% \\
\arrayrulecolor{gray!20}\midrule\arrayrulecolor{black}
  SyzDirect       & 18 & 7 & 25 & 72.0\% & 36.0\% \\
  \bottomrule
  \end{tabular}

  \vspace{14pt}

  \noindent\textbf{(b) CVE Benchmark (n=50)}
  \vspace{4pt}
  \begin{tabular}{@{}l c c c c c @{}}
  \toprule
  \textbf{Method} & \textbf{Verified} & \textbf{Mismatched} & \textbf{Missed} & \textbf{Prec.} & \textbf{Poc.} \\
\arrayrulecolor{gray!20}\midrule\arrayrulecolor{black}
  \multicolumn{6}{@{}l}{\sysname} \\
  \quad \textit{D4VS}      & 30 & 0 & 20 & 100.0\% & 60.0\% \\
  \quad \textit{Qwen3.8}   & 24 & 0 & 26 & 100.0\% & 48.0\% \\
\arrayrulecolor{gray!20}\midrule\arrayrulecolor{black}

  \multicolumn{6}{@{}l}{K-Repro} \\
  \quad \textit{D4VS}      & 21 & 7 & 22 & 75.0\% & 42.0\% \\
  \quad \textit{Qwen3.8}   & 20 & 9 & 21 & 69.0\% & 40.0\% \\
\arrayrulecolor{gray!20}\midrule\arrayrulecolor{black}
  SyzDirect       & 11 & 4 & 35 & 73.3\% & 22.0\% \\
  \bottomrule
  \end{tabular}

  \vspace{1pt}
  \raggedright\footnotesize
  \textbf{Prec.}~(Precision) $=$ Verified $/$ (Verified $+$ Mismatched).
  \textbf{Poc.}~(Poc Yield) $=$ Verified $/$ Total.
  D4VS and Qwen3.8 denote D4VS and Qwen3.8-Flash backbones, respectively.
  Both LLM methods use the same 50 cases per benchmark.
  \textbf{SyzDirect}~\cite{tan2023syzdirect} is directed greybox fuzzing (non-LLM),
  given the crash site as target and a budget of 24 hours per bug.
\end{table}

\Cref{table:rq2a} reports on the PoC generation results on both benchmarks. On Syzbot, \sysname generates PoCs successfully for 34 of the 50 target bugs on DeepSeek-V4-Flash, achieving 68.0\% PoC yield and 89.5\% precision; four additional cases trigger different kernel faults, while 12 produce no crash.
Results on Qwen-3.8-Flash have 9 mismatches and 13 missed crashes.
On the CVE benchmark, \sysname produces PoCs for 30 of the 50 bugs on DeepSeek-V4-Flash, achieving 60.0\% PoC yield and 100.0\% precision.
The results of Qwen-3.8-Flash achieved 24 verified crashes.
All crashes generated demonstrate the intended defects, indicating that \sysname can provide reliable executable evidence across diverse historical vulnerabilities.

\begin{table*}[t]
\centering
\caption{Ablation on 100 Sashiko AI Reviews.}
\label{table:ablation}
\small
\setlength{\tabcolsep}{7pt}
\begin{tabular}{@{}lcccccc@{}}
\toprule
& \textbf{Success} & \textbf{Success} & \textbf{Avg.\ Time} & \textbf{Avg.\ Cost} & \textbf{Avg.\ Tokens} & \textbf{Avg.\ Steps} \\
\textbf{Method} & \textbf{@Crash (\%)} & \textbf{@Match (\%)} & \textbf{(min)} & \textbf{(\$)} & \textbf{(input/output)} & \\
\midrule
\sysname  & \textbf{52.0}~(\textit{52/100}) & \textbf{49.0}~(\textit{49/100}) & 72.8 & \textbf{0.210} & \textbf{13.0M} / 274.8K & \textbf{136.0} \\
~~\textit{-- Matched~(49)} &  - &  - & 62.7 & 0.156 & 8.9M / 234.1K & 111.0 \\
~~\textit{-- Unmatched~(51)}     &  - &  - & 82.6 & 0.262 & 17.0M / 313.9K & 160.0 \\
\arrayrulecolor{gray!20}\midrule\arrayrulecolor{black}

\sysname$_{\text{w/o FSM}}$ & 19.0~(\textit{19/100}) & 14.0~(\textit{14/100}) & 85.0 & 0.251 & 16.1M / 310.2K & 250 \\
~~\textit{-- successful~(19)} &  - &  - & 67.9 & 0.170 & 10.0M / 243K & 152 \\
~~\textit{-- failed~(81)}     &  - &  - & 89.0 & 0.270 & 17.5M / 326K & 273 \\
\arrayrulecolor{gray!20}\midrule\arrayrulecolor{black}

\sysname$_{\text{review}}$ & 27.0~(\textit{27/100}) & 21.0~(\textit{21/100}) & 92.0 & 0.270 & 17.1M / 339.8K & 290 \\
~~\textit{-- successful~(27)} &  - &  - & 73.1 & 0.179 & 10.5M / 258K & 171 \\
~~\textit{-- failed~(73)}     &  - &  - & 99.0 & 0.303 & 19.5M / 370K & 334 \\
\arrayrulecolor{gray!20}\midrule\arrayrulecolor{black}

\sysname$_{\text{w/o E}}$ & 15.0~(\textit{15/100}) & 11.0~(\textit{11/100}) & 117.7 & 0.303 & 20.6M / 330.8K & 389 \\
~~\textit{-- successful~(15)} &  - &  - & 51.0 & 0.139 & 9.5M / 148K & 170 \\
~~\textit{-- failed~(85)}     &  - &  - & 129.5 & 0.332 & 22.5M / 363K & 428 \\
\arrayrulecolor{gray!20}\midrule\arrayrulecolor{black}

\sysname$_{\text{flat}}$ & 42.0~(\textit{42/100}) & 20.0~(\textit{20/100}) & 123.2 & 0.327 & 22.8M / 333K & 380 \\
~~\textit{-- successful~(42)} &  - &  - & 105.3 & 0.279 & 19.5M / 284.6K & 324.8 \\
~~\textit{-- failed~(58)}     &  - &  - & 136.2 & 0.361 & 25.2M / 368.1K & 420.0 \\
\arrayrulecolor{gray!20}\midrule\arrayrulecolor{black}

Claude Code & 24.0~(\textit{24/100}) & 13.0~(\textit{13/100}) & \textbf{62.4} & 0.217 & 18.2M/\textbf{107K} & 173 \\
~~\textit{-- successful~(24)} &  - &  - & 44.0 & 0.137 & 11.0M/85K & 108 \\
~~\textit{-- failed~(76)}     &  - &  - & 68.2 & 0.242 & 20.5M/114K & 194 \\

\bottomrule
\multicolumn{7}{@{}p{0.93\textwidth}@{}}{
\textbf{@Crash} denotes that a kernel crash was successfully triggered, while
\textbf{@Match} signifies that the triggered crash call trace aligns with the bug description provided in the review.
For full \sysname, Matched and Unmatched are defined by @Match, as in \Cref{table:rq1a}.
$\sysname_{w/o FSM}$ omits induced reproduction FSMs; $\sysname_{\text{w/o E}}$ excludes environment-related tools and uses a static default x86 environment; $\sysname_{flat}$ is a single agent with the full environment toolset with out environment agent; $\sysname_{review}$ includes only subsystem review knowledge; and Claude Code runs without kernel-specific knowledge or a specialized harness, with full access to the experimental environment and the same base LLM (deepseek-v4-flash).
}
\end{tabular}
\end{table*}
\vspace{-2ex}

K-Repro generates PoCs successfully for 23 of the 50 Syzbot bugs and 21 of the 50 CVEs on DeepSeek-V4-Flash, and 26 Syzbot bugs and 20 CVEs on Qwen-3.8-Flash, confirming the feasibility of agentic kernel bug reproduction.  Under the same evaluation, \sysname produces PoCs for 34 and 30 cases on DeepSeek-V4-Flash and 28 and 24 cases on Qwen-3.8-Flash, respectively. K-Repro also triggers 15 different faults on Syzbot and seven on CVEs, resulting in precisions of 60.5\% and 75.0\% and PoC Yield of 46.0\% and 42.0\% on DeepSeek-V4-Flash.
In historical bugs, K-Repro performs better than in recent AI reviews, but is still much worse than \sysname.
These results suggest that execution-grounded reproduction FSMs and adaptive environment construction can further help an agent focus its search on the intended defect; we examine the contributions of these mechanisms separately in RQ3.

SyzDirect generates PoCs successfully for 18 of the 50 Syzbot bugs and 11 of the 50 CVEs within its 24-hour budget per case, showing that directed fuzzing can reproduce a meaningful subset of known bugs without LLM-based reasoning. Its exploration, however, depends on path and state-space coverage as well as available Syzlang descriptions, making complex subsystem-specific interactions, such as BPF- and kfunc-heavy paths, harder to realize. SyzDirect also triggers different faults in seven Syzbot and four CVE cases. These results suggest that directed fuzzing and review-conditioned agentic construction offer complementary strengths.

The modest difference between Syzbot~(68\%) and CVE~(60\%) may partly reflect the characteristics of the two benchmarks. Syzbot reports typically provide structured crash diagnostics and concrete call traces, whereas CVE descriptions are often less structured and span more heterogeneous kernel versions, increasing the difficulty of both trigger reconstruction and environment setup. Despite these differences, \sysname achieves 89.5\% and 100\% precision, indicating that when it produces a crash, the resulting PoC usually demonstrates the intended fault.

\vspace{-1ex}
\begin{tcolorbox}[boxrule={0.03cm}, boxsep={0.01cm}]
\textbf{RQ2 Summary:} On 100 known kernel bugs, \sysname consistently outperforms K-Repro and SyzDirect across both benchmarks, which demonstrates the effectiveness of \sysname in generating PoCs for bugs.
\end{tcolorbox}

\vspace{-1ex}
\subsection{RQ3: Ablation Study}
\vspace{-2ex}

\begin{figure*}[t]
    \centering
    \includegraphics[width=1\textwidth]{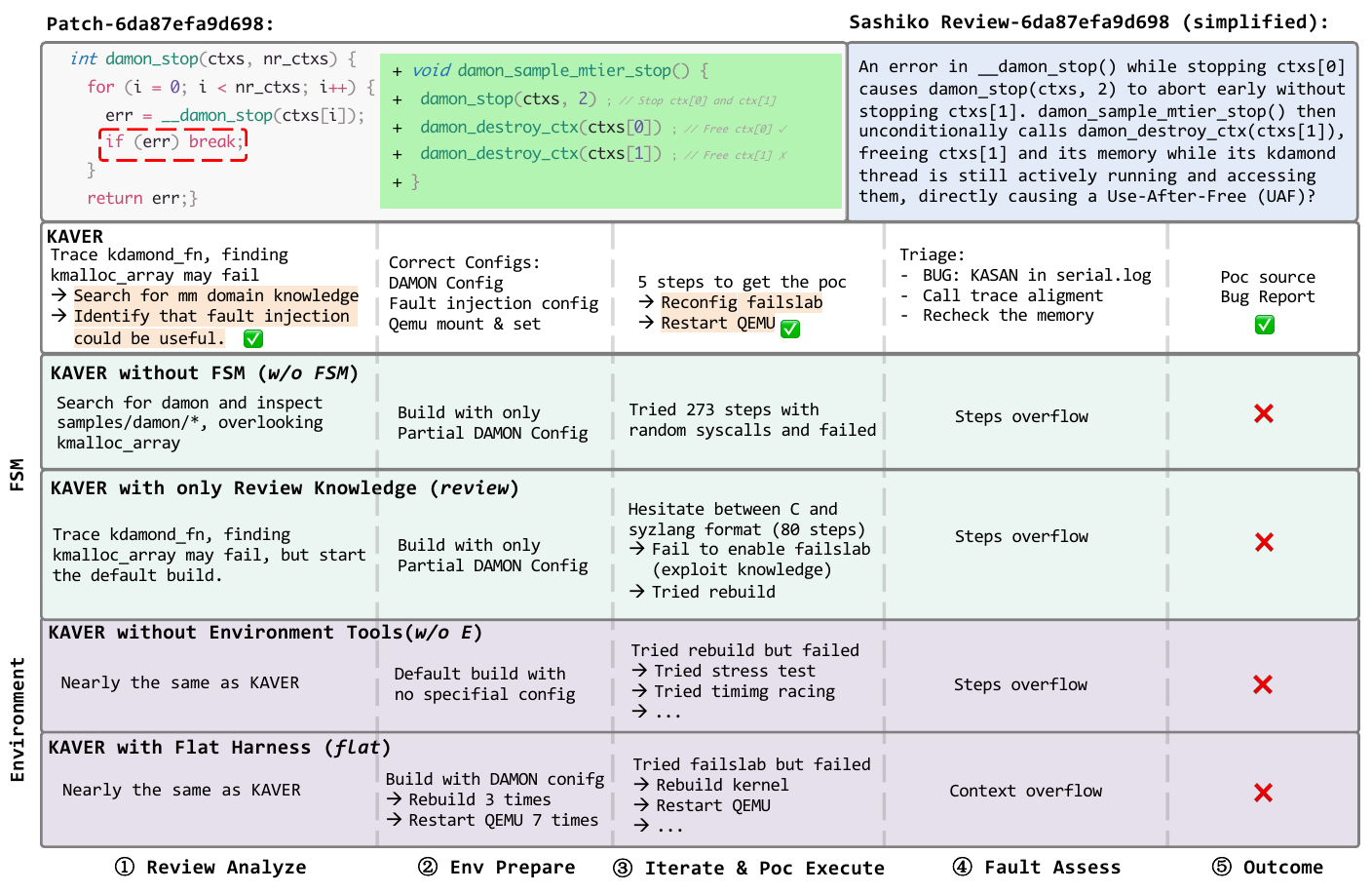}
    \caption{An ablation example showing the PoC generation performance of different \sysname variants on a memory-related bug. Full \sysname combines distilled FSM, which contains \codeztext{failslab} trigger knowledge, with governed environment reconfiguration to produce a call-trace-matching KASAN crash. }
    \label{fig:ablation}
\end{figure*}

To understand how \sysname's core mechanisms affect PoC generation for AI review, we evaluated four variants on the same 100 AI reviews on DeepSeek-V4-Flash. Two variants examine the contribution of FSMs distilled from the offline phase, while the other two examine the environment adaptation and its governance. We report their PoC generation effectiveness and execution costs alongside full \sysname and Claude Code in \Cref{table:ablation}, and use the representative example in \Cref{fig:ablation} to illustrate how these design choices affect trigger construction and environment realization.
\subsubsection{Kernel Reproduction FSMs}

$\sysname_{\text{w/o FSM}}$ removes the induced FSMs while maintaining the complete two-agent framework. Its PoC generation rate decreases from 49.0\% to 14.0\%, a 35\% drop. Without guidance from successful historical trajectories, the PoC Agent must derive the trigger primitives, kernel-state transitions, and execution realization alone, substantially enlarging the search space.

$\sysname_{\text{review}}$ retains only kernel review knowledge~\cite{bpf-agent} as reference documentation added to the system prompt of the two agents, including subsystem-specific bug patterns and code-level failure cues, while removing all of the FSM guidance. It produces PoCs that demonstrate the reviewed failures in 21 cases, compared with 14 for $\sysname_{\text{w/o FSM}}$ and 49 for the full system.
This result suggests that review knowledge helps the agent understand where and why a defect may occur, but does not by itself specify how to exploit it. As shown in the middle of \Cref{fig:ablation}, the agent identifies the possible \codeztext{kmalloc\_array} failure but does not connect it to a \codeztext{failslab} trigger, resulting in a partially configured build and repeated switching between Syzlang and C PoC formats until its step budget is exhausted.

\subsubsection{Environment Adaptation}

$\sysname_{\text{w/o E}}$ retains the induced reproduction FSMs but removes environment-related tools, leaving a static initial kernel and QEMU. It produces PoCs demonstrating the reviewed failures in 11 cases, compared with 49 for the full system, while its average step count increases from 136.0 to 389. As shown in \Cref{fig:ablation}, the agent correctly selects the \codeztext{failslab} trigger, but the default kernel lacks the required fault-injection configuration. Being unable to adapt the kernel configs, it explores infeasible stress and timing alternatives until its step budget is exhausted.

$\sysname_{\text{flat}}$ retains the FSMs and the full set of environment tools, but assigns the construction of the PoC and the adaptation of the environment to a single agent~(without an Environment agent). It produces PoCs demonstrating the reviewed failures in 20 cases, compared with 49 for full \sysname, while requiring 380.0 steps, 123.2 minutes, and \$0.327 per case on average.
These results indicate that environment mutability alone is insufficient: effective co-adaptation also requires coordinating environment changes and maintaining consistent reasoning across rebuilds.

\vspace{-1ex}
\subsubsection{Complementary Roles of the Two Mechanisms}
The ablations show that FSMs and environment adaptation address different but coupled aspects of PoC generation. Removing FSMs reduces the number of successful PoCs from 49 to 14, as the agent lacks guidance for constructing the required triggers and execution paths. Removing environment adaptation reduces the number to 11, as even a correctly identified trigger may not be executable in the initial environment. The flat harness recovers 20 successful PoCs by allowing environment changes, but still requires 380.0 steps and \$0.327 per case, indicating that environment changes alone do not provide effective coordination between PoC refinement and kernel configuration. Overall, FSMs guide the agent in constructing the execution needed to reproduce the bug, while environment adaptation makes that execution feasible in the current environment.

\vspace{-1ex}
\subsubsection{General-Purpose Agent}
We run Claude Code with the same base model and environment tools, but without induced FSMs. Claude Code triggers 24 crashes and produces PoCs demonstrating the reviewed failures in 13 cases, compared with 52 crashes and 49 PoCs for \sysname. Although its average runtime is shorter~(62.4 vs.\ 72.8 minutes), it uses more steps~(173 vs.\ 136.0), input tokens~(18.2M vs.\ 13.0M), and cost per case~(\$0.217 vs.\ \$0.210). This is \$1.67 per successful PoC, compared with \$0.43 for \sysname.

We also observe that eight of its 24 crashes result from modifying the kernel source or injecting into the target process with kernel privileges, rather than constructing a user-space trigger. Such executions do not satisfy our PoC criterion, although they demonstrate the agent's ability to reach a crash through alternative means. Overall, general-purpose agents can generate executable PoCs for some reviews, but their search efficiency remains limited.

\subsubsection{Illustrative Example}

\Cref{fig:ablation} shows an example with a use-after-free bug flagged by Sashiko. The patch stops two monitoring contexts and then destroys both, but \codeztext{damon\_stop} returns after the first error. The second context may therefore remain active when \codeztext{freed} is called, allowing its \codeztext{kdamond} thread to access released memory.
The review explains the bug but not how to trigger it. \sysname identifies a \codeztext{kmalloc\_array} failure as a way to terminate the first \codeztext{kdamond} and maps it to a \codeztext{failslab}-based trigger. The Environment Agent enables DAMON and fault injection, while the PoC Agent tunes the trigger until KASAN reports the reviewed use-after-free.

This shows why both FSM and environment adaptation are needed: $\sysname_{\text{w/o FSM}}$ misses the allocation-failure path, exploring unrelated triggers; $\sysname_{\text{review}}$ sees \codeztext{kmalloc\_array} may fail but not that \codeztext{failslab} can trigger it; $\sysname_{\text{w/o E}}$ selects \codeztext{failslab} but its static kernel lacks the required fault-injection options; $\sysname_{\text{flat}}$ can rebuild the kernel but repeatedly changes the environment and retains old-build results. Full \sysname uses the FSM to find the \codeztext{failslab} trigger, enables the required options, rebuilds the kernel, and successfully reproduces the bug.
\begin{tcolorbox}[boxrule={0.03cm}, boxsep={0.01cm}]
\textbf{RQ3 Summary:} FSMs and two-agent environment adaptation both matter. A single-agent framework produces 20 at a much higher cost. FSMs guide bug triggering; the two agents manage the environment.
\end{tcolorbox}

\section{Discussion}
\vspace{-1ex}
\textit{Implications for Generating PoCs for AI Review.}
Our findings suggest that a capable AI review should not stop at explaining why bugs may exist, but should also search for an execution that demonstrates it. \sysname shows that this review-to-PoC gap can be narrowed by combining reproduction finite-state-machines with governed multi-agent framework. On recent kernel reviews, the resulting generated PoCs indicate that an executable PoC program makes AI findings easier to assess and act upon. Rather than serving as a complete verifier, a reproducing PoC provides strong executable support for a bug report.
These results suggest that future AI review systems should provide runnable PoCs and the environments needed to execute them.

\vspace{-1ex}
\subsection{Threats to Validity}
\vspace{-1ex}

\noindent\textbf{Internal Validity.}
Several authors manually compare each crash with its bug report, but errors remain possible. Learned FSMs depend on our classification of replay steps and may omit rare but useful steps. All agent-based methods use the same model, budgets, and host environment.

\smallskip
\noindent\textbf{Construct Validity.}
We use crash traces to check whether PoCs trigger reported bugs, but may accept similar crashes or reject valid PoCs reaching the same bug through another path. Failure to generate a PoC does not imply that a review is wrong. The maintainer study is observational, as \sysname submits only reports with generated PoCs.

\smallskip
\noindent\textbf{External Validity.}
Our evaluation uses Sashiko reviews, Syzbot and CVE bugs, two base models. Results may differ with other review systems or models. Recent reviews reduce but cannot eliminate memorization risk. We mainly evaluate runtime bugs that crash the kernel; bugs without crashes or requiring specific hardware may need different PoCs and environments. FSMs learned from successful trajectories may cover common bug patterns better than rare ones.

\vspace{-1ex}
\subsection{Improvements and Future Work}
\vspace{-1ex}
We plan to improve \sysname in four directions. First, hardware-dependent bugs remain difficult because QEMU does not emulate every device; richer device models or real-hardware testbeds could support more bugs. Second, \sysname mainly handles bugs that cause observable kernel crashes; supporting silent logic errors, data corruption, and performance regressions will require different runtime checks. Third, the subsystem FSMs are learned offline and remain fixed during online PoC generation; we plan to update them from newly successful reproductions and learn more specific FSMs for different bug classes. Finally, matching a generated crash to the review still requires manual checking; a more accurate automated checker could reduce this effort and support larger deployments.

\vspace{-1ex}
\section{Related Work}
\vspace{-1ex}
\textbf{Agentic Code Review.} AI agents now review code changes at production scale. In the Linux kernel, Google's Sashiko~\cite{sashiko} runs an agentic review protocol over patches, and the bpf-next review bot~\cite{bpf-agent} automatically reviews BPF patch series in CI. Beyond kernels, agentic review is spreading across software engineering: Archer~\cite{archer} extends agentic review to compiler optimizations, where update frequency outpaces expert-review capacity, and a large-scale study of CodeRabbit~\cite{coderabbit} mines developer feedback to assess whether agentic reviews are genuinely helpful. These systems surface suspected defects but stop at identification; \sysname closes the loop by verifying a review with an executable, call-trace-matching PoC.

\textbf{Proof-of-Concept and Exploit Generation.} Automatic exploit generation began with symbolic execution over user-space programs. AEG~\cite{aeg} generated end-to-end control-flow-hijack exploits, and Mayhem~\cite{mayhem} extended this approach to stripped binaries. For the kernel, prior work turns a known crash into an exploit primitive. FUZE~\cite{FUZE} targets use-after-free, KOOBE~\cite{KOOBE} targets out-of-bounds writes, and KEPLER~\cite{kepler} turns a control-flow hijack into a stable primitive. GREBE~\cite{grebe} explores alternative error behaviors, and SyzScope~\cite{zou2022syzscope} identifies the high-risk impact of fuzzer-exposed bugs. These methods require an existing crashing input and mostly use manually encoded strategies. Recent LLM systems include K-Repro~\cite{krepro} for reproducing kernel bugs from fix commits, a program-analysis-guided agent for user-space PoCs~\cite{desai2026program}, and LLM agents that exploit memory bugs~\cite{bigsleep}. However, they rely on a generic harness and a general success criterion. \sysname instead combines distilled exploit knowledge with call-trace validation.

\textbf{Kernel Testing.} Kernel testing uses static and dynamic analysis. Dr.Checker~\cite{drchecker} and K-Miner~\cite{kminer} detect memory-corruption bugs without execution, and recent work~\cite{li2025towards} improves taint-style detection. Static analysis, however, cannot show that a bug is triggerable.
Coverage-guided fuzzing is the main dynamic approach. syzkaller~\cite{syzkaller2024} generates system-call programs from syzlang specifications. MoonShine~\cite{moonshine}, HEALER~\cite{healer}, and StateFuzz~\cite{statefuzz} improve this process by reusing execution traces, learning system-call relations, and tracking driver state, respectively. Directed fuzzers, such as SyzDirect~\cite{tan2023syzdirect}, guide execution toward a kernel location, but still seek a crash. These fuzzers are limited by syzlang coverage and do not reproduce the specific defect in a review. \sysname instead constructs a PoC guided by FSMs and validates it against the expected call trace.

\vspace{-1ex}
\section{Conclusion}
\vspace{-1ex}

We presented \sysname, a multi-agent framework that generates PoCs for AI kernel reviews by replaying historical bugs and learning one reproduction FSM per kernel subsystem. It generated PoCs for 49 of 100 recent AI reviews and 64 of 100 known bugs; maintainers replied to 35 of the 49 submitted reports and confirmed 34. Ablations show that both reproduction FSMs and environment adaptation are needed for reliable PoC generation, and that subsystem-specific FSMs plus a structured two-agent workflow can turn AI reviews into runnable PoCs—supporting kernel maintenance and helping maintainers assess AI review accuracy—thereby moving AI-assisted kernel review from textual findings to reproducible evidence.

\bibliographystyle{IEEEtran}
\bibliography{references}

\section*{Appendix A. Ethical Considerations}

\sysname generates executable PoCs for Linux kernel defects and therefore has potential dual-use implications. While such PoCs can help validate AI-generated reviews and provide concrete evidence of whether a reported defect is reachable, the same capability could lower the effort required to reproduce security-relevant kernel bugs.

We mitigate this risk through the scope and execution setting of \sysname. First, \sysname is designed for bug reproduction rather than exploitation. Its objective is to construct a user-space trigger that causes a kernel fault whose call trace matches the defect described in the review or bug report. \sysname does not attempt to transform the reproduced fault into an exploitation primitive, obtain privilege escalation, achieve arbitrary code execution, or otherwise assess post-crash exploitability. This distinction is also reflected in our evaluation criterion, where a generated PoC is considered successful only when the observed kernel fault matches the reported defect.

Second, all generated PoCs are executed in controlled and isolated test environments. \sysname builds target kernel revisions and runs them inside QEMU virtual machines with KASAN instrumentation, rather than executing generated programs on production systems. Environment adaptation, including changes to kernel configurations, emulated devices, architectures, and userspace utilities, is confined to these experimental virtual machines. Consequently, potentially destructive kernel crashes produced during PoC generation affect only disposable test instances.

Finally, \sysname operates on kernel patches, AI reviews, and historical bug reports for the purpose of determining whether a reported failure can be concretely reproduced. The framework does not autonomously target deployed hosts, scan external systems for vulnerable kernels, or provide mechanisms for delivering generated PoCs to remote machines. These design and evaluation choices constrain \sysname to controlled vulnerability validation, while preserving its intended benefit of providing executable evidence for otherwise speculative AI-generated kernel reviews.

\end{document}